# Using Platt's scaling for calibration after undersampling – limitations and how to address them

Nathan Phelps[1,*], Daniel J. Lizotte[2,3,†], and Douglas G. Woolford[1,†]

[*]Corresponding author

[†]Equal contribution

[1]Department of Statistical and Actuarial Sciences, University of Western Ontario, London, Ontario, Canada

[2]Department of Computer Science, University of Western Ontario, London, Ontario, Canada

[3]Department of Epidemiology and Biostatistics, University of Western Ontario, London, Ontario, Canada

*Email: nphelps3@uwo.ca; Postal Address: 1151 Richmond St., London, ON, N6A 3K7

ORCID for Nathan Phelps: 0000-0002-3173-3368

ORCID for Daniel J. Lizotte: 0000-0002-9258-8619

**Abstract:** When modelling data where the response is dichotomous and highly imbalanced, response-based sampling where a subset of the majority class is retained (i.e., undersampling) is often used to create more balanced training datasets prior to modelling. However, the models fit to this undersampled data, which we refer to as base models, generate predictions that are severely biased. There are several calibration methods that can be used to combat this bias, one of which is Platt's scaling. Here, a logistic regression model is used to model the relationship between the base model's original predictions and the response. Despite its popularity for calibrating models after undersampling, Platt's scaling was not designed for this purpose. Our work presents what we believe is the first detailed study focused on the validity of using Platt's scaling to calibrate models after undersampling. We show analytically, as well as via a simulation study and a case study, that Platt's scaling should not be used for calibration after undersampling without critical thought. If Platt's scaling would have been able to successfully calibrate the base model had it been trained on the entire dataset (i.e., without undersampling), then Platt's scaling might be appropriate for calibration after undersampling. If this is not the case, we recommend a modified version of Platt's scaling that fits a logistic generalized additive model to the logit of the base model's predictions, as it is both theoretically motivated and performed well across the settings considered in our study.

**Statements and declarations:** The authors do not have any relevant financial or non-financial interests to disclose.

**Funding:** We acknowledge the support of the Natural Sciences and Engineering Research Council of Canada (NSERC) through its Postgraduate Scholarship program, Discovery Grant program, and Strategic Networks program.

# 1. Introduction

Highly imbalanced, binary classification[1] problems are ubiquitous in today's world of data analytics. These problems appear in fields as varied as finance (e.g., fraud detection; Varmedja et al., 2019), health care (e.g., disease modelling; Shin et al., 2023), and wildfire (e.g., fire occurrence prediction; Phelps and Woolford, 2021b). In our study, we will assume that the minority class represents the occurrence of the outcome of interest (denoted by 1; also called the positive class), such as a fraudulent transaction, and the majority class represents its non-occurrence (denoted by 0; also called the negative class). Oftentimes, the datasets associated with imbalanced classification problems are very large (i.e., millions—or more—of observations with many covariates). Both the imbalance and size of these datasets can present modelling challenges. Fitting complex models to massive datasets can be very time-consuming and, in addition, studies have shown that some models tend to neglect the minority class in the face of substantial class imbalance (e.g., Japkowicz and Stephen, 2002). A common method for handling both these issues is undersampling, or downsampling (e.g., Wallace and Dahabreh, 2013; Moreau et al., 2020; Peng et al., 2020; Phelps and Woolford, 2021b; Burmeister et al., 2023; Shin et al., 2023).

When undersampling, all observations from the minority class are kept but only a random subset of the majority class is retained for modelling. The problem with this sampling procedure is that it biases the model. Consider a data distribution $f_{X,Y}(x, y)$, where $x$ is a vector of covariates and $y$ is a binary outcome, and a model $h$ that produces estimates $h(x) \approx \mathbb{P}(Y = 1|X = x)$. We define a perfectly calibrated model as one that generates estimates such that for any $\hat{p} = h(x)$ that can be produced by the model, $\mathbb{P}(Y = 1|h(X) = \hat{p}) = \hat{p}$, where the probability is computed over the data distribution $f_{X,Y}$. For example, for observations where the model assigns $\hat{p} = 0.3$, we expect that 30% are truly positives. Because the distribution of the training data now differs from that of new data, a model $h$ that was well-calibrated on its undersampled dataset will not be well-calibrated when used to make predictions on new data. Generally, a model trained on undersampled data will output estimates that overestimate the true outcome probabilities. This bias induced by undersampling is a serious issue because having poorly calibrated probability estimates can hinder the effectiveness of the model for use in practice, altering prevalence estimates and potentially leading to suboptimal decision-making (e.g., Phelps and Woolford, 2021a; Guilbert et al., 2024). Consequently, it is important to adjust the probability estimates of the model to try to obtain well-calibrated estimates whenever possible.

There are several different methods for calibrating models after undersampling. One of the most common approaches is Platt's scaling (Platt, 1999). Despite its popularity in this situation (e.g., Wallace and Dahabreh, 2013; Moreau et al., 2020; Peng et al., 2020; Phelps and Woolford, 2021b; Burmeister et al., 2023; Shin et al., 2023), Platt's scaling was originally designed for another purpose – augmenting the output of support vector machines to obtain calibrated probabilities. It has since been used to calibrate models in other situations, sometimes because models learned from undersampled data (e.g., Wallace and Dahabreh, 2013; Moreau et al., 2020; Peng et al., 2020; Phelps and Woolford, 2021b; Burmeister et al., 2023; Shin et al., 2023) and sometimes because models were miscalibrated for other reasons (e.g., Guo et al., 2017; Ojeda et al., 2023). Platt's scaling involves fitting a logistic regression model, enforcing a sigmoid relationship between covariates and probabilities. Because of this restriction, the validity of using Platt's scaling outside of its original purpose has been debated; some have criticized it (e.g., Naeini et al., 2015; Kull et al., 2017), but Böken (2021) showed that it is justifiable to use Platt's scaling in more scenarios than other works have suggested. From what we have seen in the literature involving undersampling, it does not appear that the appropriateness of using Platt's scaling in the context of calibration after undersampling is well-understood.

---

[1] Throughout this paper, we use "classification" to refer to situations where the observation is categorical (and usually, binary.) This does not mean our model must be a "classifier" in the sense that it outputs a class prediction; the model may output class probabilities, but we still consider this classification.

Our work presents what we believe is the first detailed study on the validity of using Platt's scaling to calibrate models after undersampling. First, we analytically show that Platt's scaling is incapable of properly calibrating a model perfectly fit to an undersampled dataset, leading to incorrect estimates of conditional probabilities. We also show how Platt's scaling can be modified so that it can provide correct estimates. Next, we demonstrate that traditional Platt's scaling can be effective when the model fit to undersampled data has a specific systematic error. Finally, we consider another adjustment for improving performance in more general settings.

In Section 2, we outline the calibration approaches considered in this study; Sections 3 and 4 provide a simulation study and an illustrative case study in the field of forestry, where we investigate a modified version of the Cover Type dataset from the UC Irvine machine learning repository; and in Section 5, we provide conclusions and practical recommendations.

## 2. Calibration after undersampling

We consider calibrating binary classification models fit to undersampled data. In this context, we have a dichotomous response $Y$ that takes values 0 (i.e., the negative class) or 1 (i.e., the positive class) and associated covariates $X$, and our goal is to estimate $p(x) = \mathbb{P}(Y = 1|X = x) \in (0,1)$. We also introduce a dichotomous variable $S$ that indicates which observations will be used for estimating $p(x)$ by first estimating $\gamma(x) = \mathbb{P}(Y = 1|X = x, S = 1) \in (0,1)$ and then transforming $\gamma$ using a calibration function $\kappa$, giving $p(x) = \kappa(\gamma(x))$. Note that if $S$ is independent of $X$ and $Y$ (e.g., if we take a simple random sample of the training set), then $\kappa$ is the identity function. However, when undersampling, we keep all positive instances (i.e., $S = 1$ for all positive instances) and negative instances are kept only with probability $\pi_0$. We refer to the model that learns from the undersampled training dataset as the base model. If the calibration function $\kappa$ is learned using another model, we refer to this model as the calibration or secondary model. This approach is an instance of model stacking, whereby a meta learner (the calibration model) is trained based on outputs of one or more base models. The calibration function can also be analytically computed (Dal Pozzolo et al., 2015b). A benefit of analytical calibration is that it does not require learning the calibration function from another full (i.e., not undersampled) dataset, unlike the model-based approaches. However, this is not typically a problem for the model-based approaches because this data is generally available, and it is not computationally expensive to learn this function because it is only based on one covariate. In the next two subsections, we outline more pros and cons of specific calibration approaches.

**2.1 Baseline calibration methods**

Our study focuses on Platt's scaling, so the baseline methods are two other common calibration methods: analytical calibration (Dal Pozzolo et al., 2015b) and isotonic regression (e.g., Zadrozny and Elkan, 2002).

*2.1.1 Analytical calibration*

A foundational assumption of empirical modelling is that the data used to train the model follows the same distribution as future data (or testing data). Multiple studies have considered the case where this assumption is not met, developing an analytical solution to this problem (e.g., Elkan, 2001; Saerens et al., 2002; Dal Pozzolo et al., 2015b). Dal Pozzolo et al. (2015b) considered the specific case where this assumption is not met because of undersampling. Using Bayes' rule, they derived the calibration function shown in Eq. 1.

$$\kappa(\gamma) = \frac{\gamma \pi_0}{1 - \gamma + \gamma \pi_0} \qquad \text{Eq. 1}$$

To calibrate the predictions of a base model fit to an undersampled training dataset, $\hat{\gamma}$, we can use this equation but with the predictions in place of $\gamma$. Naturally, if the base model is perfect (i.e., $\hat{\gamma} = \gamma$), then this analytical approach perfectly calibrates the probability estimates. However, this approach does not account for any error in the base model. Therefore, it may struggle in practice, when models are imperfect.

*2.1.2 Isotonic regression*

Unlike the analytical method, isotonic regression (e.g., Zadrozny and Elkan, 2002) is very flexible, so it can account for error in the base model. In our case, isotonic regression aims to minimize $\sum_i [y_i - \kappa(\hat{\gamma}_i)]^2$, where $\kappa$ is a step function. This model is learned using the pair-adjacent violators (PAV) algorithm (Ayer et al., 1955), and the only restriction imposed on $\kappa$ is that it must be monotonically non-decreasing. Because $\kappa$ is a step function, the relationship between the base model's predictions and the new probability estimates is not smooth when this method is used.

**2.2 Platt's scaling and its variations**

Like isotonic regression, Platt's scaling (Platt, 1999) uses a secondary model to calibrate the predictions of the base model. Platt's scaling involves fitting a logistic regression model to the $y$'s using the $\hat{\gamma}$'s as the predictor. In the original paper, Platt (1999) slightly modified the responses to perform regularization, but Platt's scaling has often been implemented without this (e.g., Phelps and Woolford, 2021b; Ojeda et al., 2023). It is therefore sometimes called logistic calibration (e.g., Kull et al., 2017), although some still make a distinction between the two (Ojeda et al., 2023). In our work, we leave the responses as 0/1 variables. When using logistic regression, we assume that $Y \sim \text{Bernoulli}(p)$ and that there is a linear relationship between the logit of the $p$'s and the $\hat{\gamma}$'s (see Eq. 2).

$$\log\left(\frac{p}{1-p}\right) = \beta_0 + \beta_1 \hat{\gamma} \qquad Eq.\ 2$$

After fitting the logistic regression model (i.e., learning estimates, $\hat{\beta}_0$ and $\hat{\beta}_1$, for $\beta_0$ and $\beta_1$, respectively), we obtain the following calibration model from this approach:

$$\kappa(\hat{\gamma}) = \frac{\exp(\hat{\beta}_0 + \hat{\beta}_1 \hat{\gamma})}{1 + \exp(\hat{\beta}_0 + \hat{\beta}_1 \hat{\gamma})} \qquad Eq.\ 3$$

To assess the validity of Platt's scaling as a method for calibration after undersampling, we must determine if the relationship assumed by this approach is reasonable. Naturally, this depends on the base model, since it outputs the $\hat{\gamma}$'s. We first consider the case where the base model is perfect, resulting in Theorem 1.

**Theorem 1.** If the base model provides perfect estimates based on the undersampled training dataset (i.e., $\hat{\gamma} = \gamma$), Platt's scaling cannot provide properly calibrated estimates of the $p$'s.

**Proof:** Since we are considering a perfect base model, we can substitute Eq. 1 for $p$ into Eq. 2 to determine the true relationship between $p$ and $\gamma = \hat{\gamma}$ on the log odds scale. After some algebra, the result of this substitution is Eq. 4.

$$\log\left(\frac{p}{1-p}\right) = \log(\pi_0) + \log\left(\frac{\gamma}{1-\gamma}\right) \qquad Eq.\ 4$$

Clearly, the relationship between the $p$'s and the $\gamma$'s is not linear on the log odds scale. Thus, even though the base model perfectly modeled its training dataset, the secondary model that will be learned from Platt's scaling cannot possibly properly adjust the $\hat{\gamma}$'s to achieve a perfect final model. □

For those familiar with both the literature on Platt's scaling and on undersampling, the result of Theorem 1 might be expected. One of the criticisms of Platt's scaling is its inability to leave a perfect model perfectly calibrated (e.g., Kull et al., 2017), and logistic models trained on undersampled data are calibrated through only adjusting their intercept[2] (e.g., Taylor et al., 2013; Phelps and Woolford, 2021b), so it is intuitive that Platt's scaling is unable to calibrate a model that is perfect with respect to the undersampled data. However, Platt's scaling's inability to calibrate such models is more disguised in this setting, as Platt's scaling generally will still improve calibration due to the base model's extreme overprediction because of learning from the undersampled dataset (e.g., Phelps and Woolford, 2021a).

Theorem 1 shows that Platt's scaling cannot properly calibrate a perfect base model, but a corollary of Eq. 4 is that a simple transformation can be done to remedy this problem. Rather than using $\hat{\gamma}$ as the covariate, we can use $\log\{\hat{\gamma}/(1-\hat{\gamma})\}$. Although not specifically in the context of calibration after undersampling, this transformation has been considered in the calibration literature before, such as by Kull et al. (2017) for implementing beta calibration and by Leathart et al. (2017) for fitting probability calibration trees (which involve fitting a logistic regression model in the leaf nodes). Böken (2021) pointed out that Platt's scaling was designed for a predictor that can take any real value, so it makes sense to use this transformation to convert predictions from [0,1] to the real line. Via their simulation study, Ojeda et al. (2023) found that using the logit transformation generally improved calibration. We of course cannot expect the base model to be perfect in practice, but we expect that defining the logistic regression model using this transformation of $\hat{\gamma}$ will still be effective for models where $\hat{\gamma} \approx \gamma$.

---

[2] Note that for logistic models, this adjustment to the intercept is equivalent to using the analytical calibration of Dal Pozzolo et al. (2015).

Models often exhibit systematic estimation errors (e.g., Niculescu-Mizil and Caruana, 2005; Guo et al., 2017; Guilbert et al., 2024). Some models tend to push probability estimates towards 0.5, while others push estimates towards extreme values (i.e., 0 or 1). We now consider models where $\hat{\gamma}$ and $\gamma$ have a sigmoidal relationship, represented by Eq. 5.

$$\gamma = \frac{1}{1 + \exp\left[-k(\hat{\gamma} - m)\right]} \qquad Eq.\ 5$$

Here, $k$ and $m$ are arbitrary constants. Mathematically, we require only that $k \in \mathbb{R}$ and $m \in [0,1]$, but it is worth noting that settings within these bounds may not lead to a reasonable representation of a model. Under settings that reasonably represent a model (e.g., $k = 10$ and $m = 0.5$), this relationship represents a model whose estimates are pushed towards 0.5. Calibrating models that err in the form indicated by Eq. 5 is considered a valid use of Platt's scaling (e.g., Kull et al., 2017; Leathart et al., 2017) because the logistic regression model's parametric assumptions are met. In Theorem 2, we show that these assumptions are still met when the base model learns from an undersampled dataset.

**Theorem 2.** If the base model's predictions (i.e., the $\hat{\gamma}$'s), have a sigmoidal relationship with the $\gamma$'s, then Platt's scaling can provide properly calibrated estimates of the $p$'s.

**Proof:** We can substitute Eq. 5 for $\gamma$ in Eq. 4. After some algebra, we can obtain Eq. 6.

$$\log\left(\frac{p}{1-p}\right) = \log(\pi_0) - km + k\hat{\gamma} \qquad Eq.\ 6$$

The relationship between the logit of the $p$'s and the $\hat{\gamma}$'s is now linear, so the assumption of the logistic regression model is met. A logistic regression model with the learned coefficients $\hat{\beta}_0 = \log(\pi_0) - km$ and $\hat{\beta}_1 = k$ would perfectly calibrate this base model. □

We have addressed how models can be calibrated after undersampling if the model is perfect or if its predictions have a sigmoid shape with the true probabilities, but we have not addressed models with a tendency to push estimates towards extreme values or models that deviate somewhat from either perfect prediction or a perfect sigmoid shape. In both cases, we cannot derive a simple transformation that will allow us to satisfy the assumptions of logistic regression. However, this does not preclude us from being able to modify Platt's scaling to obtain better probability estimates. Rather than instituting the restrictive assumptions of logistic regression, a logistic generalized additive model (GAM) can be used. This has been done to calibrate models after undersampling in a few studies (e.g., Coussement and Buckinx, 2011; Phelps and Woolford, 2021b), but does not seem to be a common approach. Logistic GAMs use smoothers to model the relationship between covariates and the outcome, relaxing the assumption of linearity on the log odds scale so that non-linear relationships can be modeled. It is important to note that, given enough data, a logistic GAM will converge to a logistic regression model when the linearity assumption holds.

In the coming sections, we will evaluate the four variations of Platt's scaling outlined in this section: traditional Platt's scaling, Platt's scaling with the logit transformation, Platt's scaling with a logistic GAM, and Platt's scaling with both the logit transformation and a logistic GAM. Through both a simulation study and a case study, these methods will be compared to the two baseline calibration methods we have also discussed: analytical calibration and isotonic regression.

## 3. Simulation study

### 3.1 Methodology

Our simulation study was implemented in R (R Core Team, 2023).

*3.1.1 Simulating data*

To be able to compare different calibration methods, we simulated datasets with known outcome probabilities. Three different data generating processes, each with different mean probabilities, were considered. This was done so that we could study the effects of different levels of class imbalance. Each data generating process creates data that are imbalanced in terms of the response variable, making them amenable to undersampling. The mean probabilities are approximately 0.0022, 0.0208, and 0.1109. Specific details about how data were generated are given in the Appendix. We created calibration datasets with 100 000 and 1 000 000 observations so that we could investigate the effectiveness of the methods with varying amounts of data. Our testing dataset had 1 000 000 observations.

*3.1.2 Simulating model outputs*

Rather than modelling the output of our simulated datasets as a function of the covariates, we defined hypothetical base models with various estimation errors. The first hypothetical base model we considered is a perfect model (i.e., $\hat{\gamma} = \gamma$). Note that this model still required calibration to account for the undersampling process. Next, we considered models whose generated $\hat{\gamma}$'s systematically deviate from the $\gamma$'s. The model that pushes probability estimates towards 0.5 is represented by Eq. 7, while the model that pushes probability estimates towards extreme values (i.e., 0 and 1) is represented by Eq. 8. Fig. 1 shows these relationships.

$$\hat{\gamma} = min\left[max\left(\frac{-1}{10}\log\left(\frac{1}{\gamma} - 1\right) + 0.5, 0\right), 1\right] \quad\quad Eq.\ 7$$

$$\hat{\gamma} = \frac{1}{1 + \exp[-10(\gamma - 0.5)]} \qquad Eq.\ 8$$

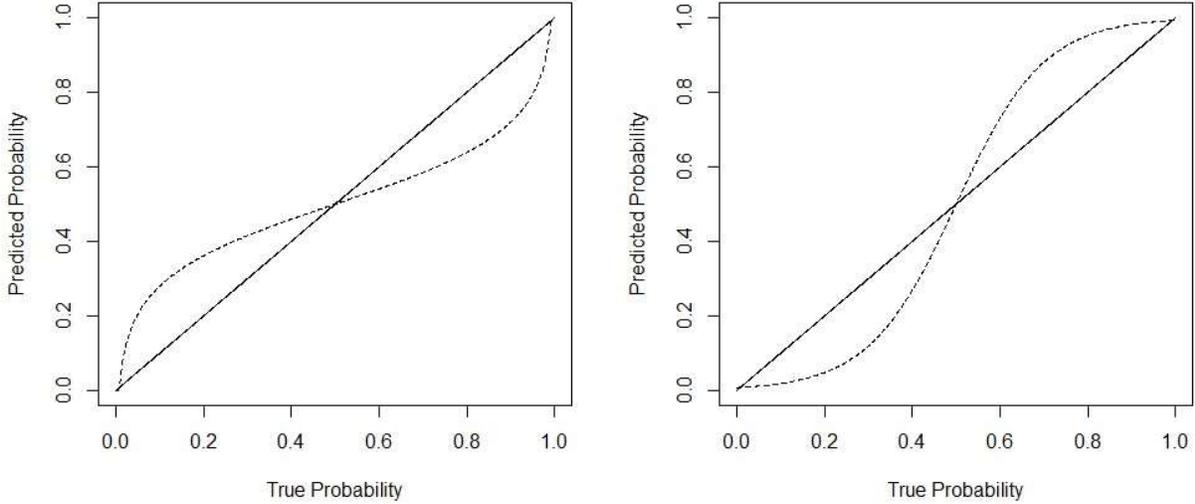

**Fig. 1** The relationship between the predicted probability and the true probability for a perfect model (solid) and a model with systematic estimation error (dashed). The left plot shows a model that pushes probabilities towards 0.5 (Eq. 7), while the right plot shows a model that pushes probabilities towards 0 and 1 (Eq. 8)

Finally, we considered a model that generates nearly perfect predictions in expectation, but does at times make larger errors. This was implemented by incorporating noise on the log odds scale of the base model. This noise was added via a Normally distributed random variable with a mean of zero and standard deviation of 0.2. Because the datasets are imbalanced, this results in $\hat{\gamma}$'s with a mean slightly larger than the mean of the $\gamma$'s. A scatterplot showing the relationship between the $\hat{\gamma}$'s and the $\gamma$'s when the mean outcome probability is 0.0208 is shown in Fig. 2.

To obtain the $\gamma$'s for each data generating process, we set sampling rates that would generate approximately balanced training datasets. For the data generating processes described in Section 3.1.1, we used sampling rates of $\pi_0 = 0.0023$, $\pi_0 = 0.02125$, and $\pi_0 = 0.125$.

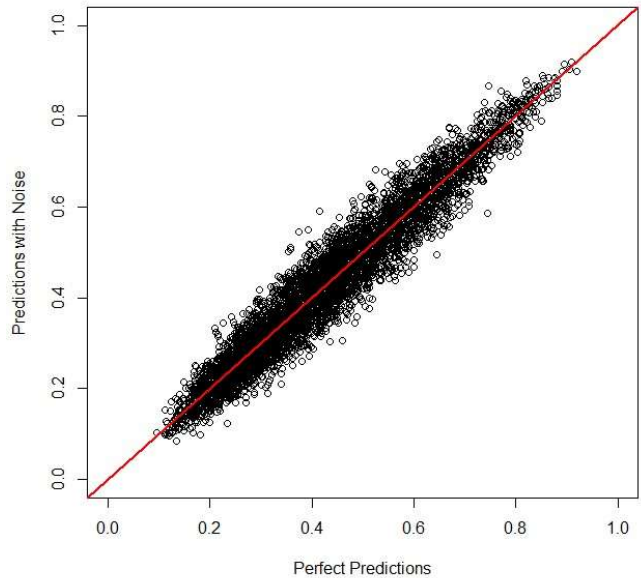

**Fig. 2** An example of the relationship between the γ̂'s and the γ's when the base model's predictions are altered by a noise variable. The scatterplot shows 5000 observations obtained from the data generating process with a mean outcome probability of 0.0208. The red line is the 45° line

*3.1.3 Implementing and evaluating the calibration methods*

To implement the four variations of Platt's scaling, we used the glm and gam functions from the stats and mgcv packages (Wood, 2011), respectively. Analytical calibration was straightforward to implement using only base R. Like Platt's scaling, isotonic regression was implemented using functions from the stats package. For each isotonic regression model, we fit the model then converted it to a step function so that we could make predictions on the testing dataset.

We evaluated the calibration methods qualitatively by plotting the probability estimates against the true probabilities. To quantitatively evaluate the methods, we measured the gap between the probability estimates and the true probabilities using root mean squared error (RMSE) and mean absolute error (MAE). We also computed Brier score (BS) and negative logarithmic score (NLS) to demonstrate how the differences between approaches might be evaluated in a setting with real data, where the true probabilities are unknown. When computing NLS, we had to modify predictions that were exactly 0 or 1 because this led to undefined values. Predictions of 0 and 1 were set to 0.00001 and 0.99999, respectively.

**3.2 Simulation study results**

*3.2.1 Perfect base model*

With a perfect base model, Platt's scaling will fall short of the analytical method no matter what variation is used. However, we were still interested in seeing how close variations of Platt's scaling can get to perfect calibration, as well as how they compare to isotonic regression. The results for the perfect base model are shown in Table 1. As expected from the analysis shown in Section 2, traditional Platt's scaling tended to perform quite poorly. However, using the logit transformation largely fixed this problem, resulting in Platt's scaling with the logit transformation performing second best of the calibration methods, behind only analytical calibration. Using this transformation but fitting a GAM also tended to perform quite well if the mean outcome success rate or size of the calibration dataset were sufficiently large.

In Fig. 3, we show the probability estimates from each calibration approach for the data generating process with a mean probability of success of approximately 0.0022 and a calibration dataset with 1 000 000 observations. This illustrates that the estimates can at times be dramatically different depending on the calibration method used. In particular, the Platt's scaling approaches without the logit transformation severely underestimated success probabilities for the cases most likely to be a success. Although this underestimation occurred for a relatively small number of instances, these are generally the cases we care about the most in the imbalanced classification setting.

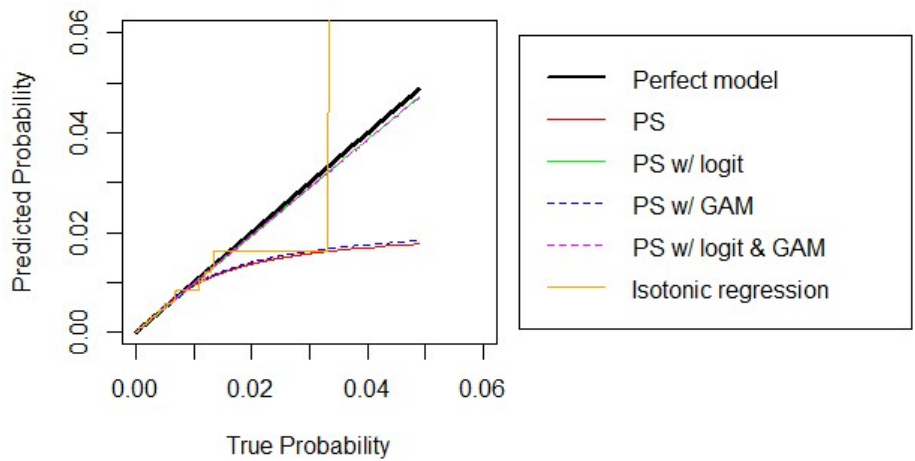

**Fig. 3** Probability estimates on the testing dataset from Platt's scaling (PS) and its variations, as well as two baseline methods. The data generating process here has a mean probability of success of approximately 0.0022. The base model is perfect and the calibration methods were trained on a dataset with 1 000 000 observations

**Table 1** The root mean squared error (RMSE), mean absolute error (MAE), Brier score (BS), and negative logarithmic score (NLS) for Platt's scaling (PS) and its variations, as well as two baseline methods. The base model used here is perfect

| Size of calibration dataset | Mean outcome success rate | Calibration method | RMSE ($\times 10^4$) | MAE ($\times 10^4$) | BS ($\times 10^3$) | NLS |
|---|---|---|---|---|---|---|
| 100 000 | 0.0022 | PS | 4.48 | 1.12 | 2.13 | 14 637 |
| | | PS w/ logit | 1.77 | 0.94 | 2.13 | 14 630 |
| | | PS w/ GAM | 4.49 | 2.69 | 2.13 | 14 655 |
| | | PS w/ logit and GAM | 4.50 | 2.82 | 2.13 | 14 657 |
| | | Analytical | 0.00 | 0.00 | 2.13 | 14 629 |
| | | Isotonic regression | 8.16 | 4.24 | 2.13 | 14 727 |
| | 0.0208 | PS | 31.09 | 10.83 | 20.25 | 95 732 |
| | | PS w/ logit | 2.05 | 1.49 | 20.24 | 95 681 |
| | | PS w/ GAM | 12.25 | 5.37 | 20.24 | 95 693 |
| | | PS w/ logit and GAM | 5.80 | 2.40 | 20.24 | 95 683 |
| | | Analytical | 0.00 | 0.00 | 20.24 | 95 679 |
| | | Isotonic regression | 42.49 | 13.58 | 20.26 | 96 031 |
| | 0.1109 | PS | 87.78 | 43.88 | 92.80 | 322 032 |
| | | PS w/ logit | 16.97 | 13.26 | 92.72 | 321 833 |
| | | PS w/ GAM | 25.76 | 14.93 | 92.73 | 321 842 |
| | | PS w/ logit and GAM | 16.97 | 13.26 | 92.72 | 321 833 |
| | | Analytical | 0.00 | 0.00 | 92.72 | 321 830 |
| | | Isotonic regression | 65.47 | 41.76 | 92.76 | 322 035 |
| 1 000 000 | 0.0022 | PS | 3.91 | 1.07 | 2.13 | 14 636 |
| | | PS w/ logit | 0.57 | 0.26 | 2.13 | 14 628 |
| | | PS w/ GAM | 3.68 | 1.00 | 2.13 | 14 635 |
| | | PS w/ logit and GAM | 0.58 | 0.26 | 2.13 | 14 628 |
| | | Analytical | 0.00 | 0.00 | 2.13 | 14 629 |
| | | Isotonic regression | 4.10 | 1.64 | 2.13 | 14 643 |
| | 0.0208 | PS | 30.76 | 11.22 | 20.25 | 95 732 |
| | | PS w/ logit | 3.41 | 1.57 | 20.24 | 95 681 |
| | | PS w/ GAM | 10.84 | 3.46 | 20.24 | 95 690 |
| | | PS w/ logit and GAM | 3.41 | 1.57 | 20.24 | 95 681 |
| | | Analytical | 0.00 | 0.00 | 20.24 | 95 679 |
| | | Isotonic regression | 15.25 | 6.64 | 20.24 | 95 705 |
| | 0.1109 | PS | 86.88 | 40.49 | 92.80 | 322 027 |
| | | PS w/ logit | 4.25 | 3.70 | 92.72 | 321 829 |
| | | PS w/ GAM | 17.09 | 7.90 | 92.72 | 321 831 |
| | | PS w/ logit and GAM | 4.25 | 3.70 | 92.72 | 321 829 |
| | | Analytical | 0.00 | 0.00 | 92.72 | 321 830 |
| | | Isotonic regression | 33.62 | 20.83 | 92.73 | 321 860 |

*3.2.2 Base model pushes probability estimates towards 0.5*

When the base model pushes probability estimates towards 0.5, the relative effectiveness of the calibration methods changed considerably (see Table 2). With this change, traditional Platt's scaling was generally the most effective approach, as expected based on the analysis in Section 2. Platt's scaling with a GAM was the next best method; like with a perfect base model, if the mean outcome success rate or size of the calibration dataset were sufficiently large, using a GAM was just as effective as traditional Platt's scaling. However, there was a considerable gap between the two approaches when the mean outcome success rate was 0.0022 and 0.0208 when only 100 000 observations were used for calibration. One unexpected result was that Platt's scaling with the logit transformation (with or without the GAM) performed best in terms of RMSE when the mean outcome success rate was 0.0022 and 100 000 observations were in the calibration dataset. This may simply be due to randomness in the simulation process, however, as traditional Platt's scaling outperformed Platt's scaling with the logit transformation in all other cases, including when the mean outcome success rate was the same but more observations were used for calibration.

In Fig. 4, we compare the calibration approaches for the data generating process with a mean success rate of 0.0208 and a calibration dataset with 100 000 observations. The difference between traditional Platt's scaling and Platt's scaling with a GAM is evident, but relatively small compared to the differences between other methods. Analytical calibration performed terribly in this setting. Isotonic regression also struggled but was still considerably better than analytical calibration.

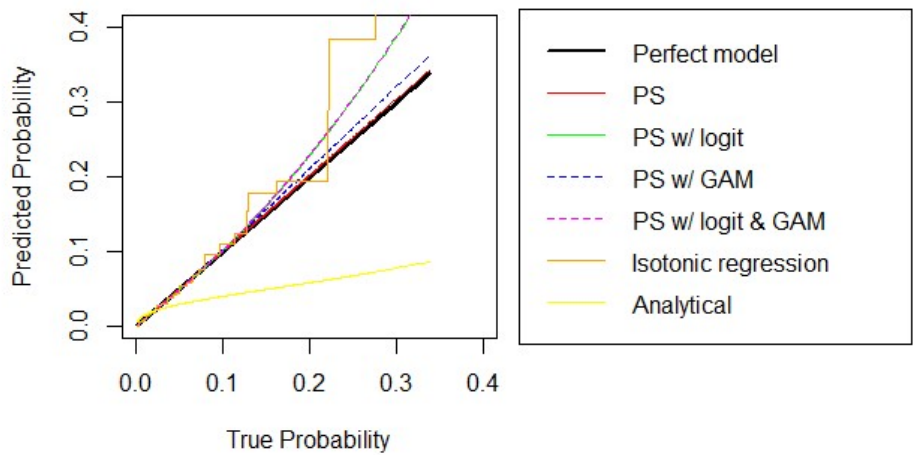

**Fig. 4** Probability estimates on the testing dataset from Platt's scaling (PS) and its variations, as well as two baseline methods. The data generating process here has a mean probability of success of approximately 0.0208. The base model pushes probability estimates towards 0.5 and the calibration methods were trained on a dataset with 100 000 observations

**Table 2** The root mean squared error (RMSE), mean absolute error (MAE), Brier score (BS), and negative logarithmic score (NLS) for Platt's scaling (PS) and its variations, as well as two baseline methods. The base model used here pushes probabilities towards 0.5

| Size of calibration dataset | Mean outcome success rate | Calibration method | RMSE ($\times 10^4$) | MAE ($\times 10^4$) | BS ($\times 10^3$) | NLS |
|---|---|---|---|---|---|---|
| 100 000 | 0.0022 | PS | 1.77 | 0.94 | 2.13 | 14 630 |
| | | PS w/ logit | 1.59 | 1.08 | 2.13 | 14 630 |
| | | PS w/ GAM | 4.50 | 2.82 | 2.13 | 14 657 |
| | | PS w/ logit and GAM | 1.59 | 1.08 | 2.13 | 14 630 |
| | | Analytical | 14.51 | 8.18 | 2.13 | 14 851 |
| | | Isotonic regression | 8.16 | 4.24 | 2.13 | 14 727 |
| | 0.0208 | PS | 2.05 | 1.49 | 20.24 | 95 681 |
| | | PS w/ logit | 11.26 | 2.51 | 20.24 | 95 687 |
| | | PS w/ GAM | 5.80 | 2.40 | 20.24 | 95 683 |
| | | PS w/ logit and GAM | 11.53 | 2.59 | 20.24 | 95 687 |
| | | Analytical | 130.10 | 75.24 | 20.41 | 97 963 |
| | | Isotonic regression | 42.49 | 13.58 | 20.26 | 96 031 |
| | 0.1109 | PS | 16.97 | 13.26 | 92.72 | 321 833 |
| | | PS w/ logit | 27.15 | 13.39 | 92.73 | 321 844 |
| | | PS w/ GAM | 16.97 | 13.26 | 92.72 | 321 833 |
| | | PS w/ logit and GAM | 25.47 | 13.49 | 92.73 | 321 842 |
| | | Analytical | 506.98 | 346.29 | 95.28 | 331 505 |
| | | Isotonic regression | 65.47 | 41.76 | 92.76 | 322 035 |
| 1 000 000 | 0.0022 | PS | 0.57 | 0.26 | 2.13 | 14 628 |
| | | PS w/ logit | 1.24 | 0.44 | 2.13 | 14 628 |
| | | PS w/ GAM | 0.58 | 0.26 | 2.13 | 14 628 |
| | | PS w/ logit and GAM | 1.69 | 0.46 | 2.13 | 14 630 |
| | | Analytical | 14.41 | 8.18 | 2.13 | 14 851 |
| | | Isotonic regression | 4.10 | 1.64 | 2.13 | 14 643 |
| | 0.0208 | PS | 3.41 | 1.57 | 20.24 | 95 681 |
| | | PS w/ logit | 11.99 | 2.01 | 20.24 | 95 686 |
| | | PS w/ GAM | 3.41 | 1.57 | 20.24 | 95 681 |
| | | PS w/ logit and GAM | 5.62 | 2.04 | 20.24 | 95 684 |
| | | Analytical | 130.10 | 75.24 | 20.41 | 97 963 |
| | | Isotonic regression | 15.25 | 6.64 | 20.24 | 95 705 |
| | 0.1109 | PS | 4.25 | 3.70 | 92.72 | 321 829 |
| | | PS w/ logit | 21.34 | 8.98 | 92.73 | 321 839 |
| | | PS w/ GAM | 4.25 | 3.70 | 92.72 | 321 829 |
| | | PS w/ logit and GAM | 9.82 | 6.61 | 92.72 | 321 829 |
| | | Analytical | 506.98 | 346.29 | 95.28 | 331 505 |
| | | Isotonic regression | 33.62 | 20.83 | 92.73 | 321 860 |

*3.2.3 Base model pushes probability estimates towards 0 or 1*

Again, the relative effectiveness of the calibration methods changed when the base model pushes probability estimates towards extreme values (see Table 3). Here, the best method was generally Platt's scaling using both the logit transformation and a GAM. This method was only beaten once, and even then, only one method – Platt's scaling with the logit transformation – was able to beat it. This occurred when the mean outcome success rate was 0.0022 and the calibration dataset had 100 000 observations. Behind Platt's scaling with both the logit transformation and a GAM, the second-best method varied. For relatively small success rates and calibration datasets, Platt's scaling with the logit transformation was second-best (or the best in the one case already discussed). When success rates and calibration datasets were larger, isotonic regression performed better.

In Fig. 5, we examine the predictions on a testing dataset from the data generating process with a mean success rate of 0.1109 and a calibration dataset with 100 000 observations. It is immediately clear that some calibration methods really struggled in this setting. Analytical calibration was very poor, underestimating the true probabilities for small probabilities and dramatically overestimating them for larger probabilities. Traditional Platt's scaling was also very poor for cases with probabilities larger than 0.2. All methods struggled for cases with the highest probabilities, but Platt's scaling with both the logit transformation and a GAM and isotonic regression provided the most reasonable estimates.

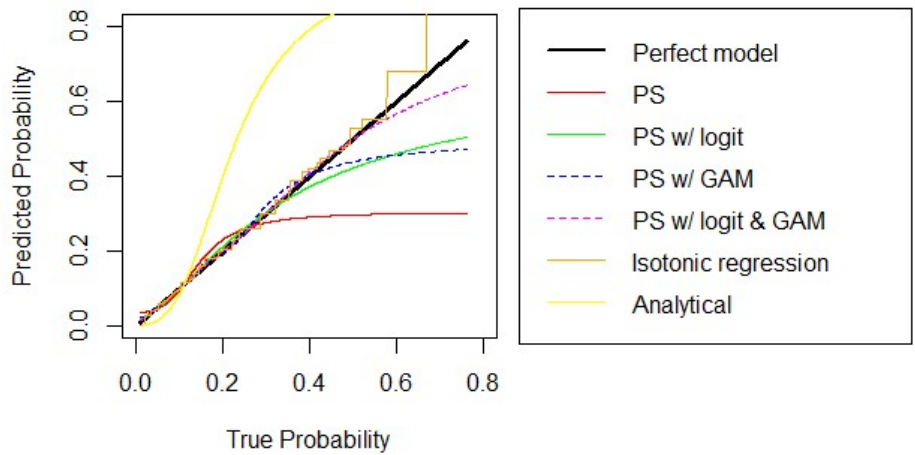

**Fig. 5** Probability estimates on the testing dataset from Platt's scaling (PS) and its variations, as well as two baseline methods. The data generating process here has a mean probability of success of approximately 0.1109. The base model pushes probability estimates towards 0 or 1 and the calibration methods were trained on a dataset with 100 000 observations

**Table 3** The root mean squared error (RMSE), mean absolute error (MAE), Brier score (BS), and negative logarithmic score (NLS) Platt's scaling (PS) and its variations, as well as two baseline methods. The base model used here pushes probabilities towards 0 or 1

| Size of calibration dataset | Mean outcome success rate | Calibration method | RMSE ($\times 10^4$) | MAE ($\times 10^4$) | BS ($\times 10^3$) | NLS |
|---|---|---|---|---|---|---|
| 100 000 | 0.0022 | PS | 8.82 | 3.27 | 2.13 | 14 680 |
| | | PS w/ logit | 4.48 | 1.17 | 2.13 | 14 637 |
| | | PS w/ GAM | 6.81 | 3.64 | 2.13 | 14 681 |
| | | PS w/ logit and GAM | 4.49 | 2.69 | 2.13 | 14 655 |
| | | Analytical | 88.13 | 29.67 | 2.21 | 16 266 |
| | | Isotonic regression | 8.16 | 4.24 | 2.13 | 14 727 |
| | 0.0208 | PS | 75.23 | 32.16 | 20.30 | 96 128 |
| | | PS w/ logit | 31.09 | 10.83 | 20.25 | 95 732 |
| | | PS w/ GAM | 32.42 | 11.74 | 20.25 | 95 753 |
| | | PS w/ logit and GAM | 12.25 | 5.37 | 20.24 | 95 693 |
| | | Analytical | 569.12 | 238.58 | 23.48 | 108 924 |
| | | Isotonic regression | 42.49 | 13.58 | 20.26 | 96 031 |
| | 0.1109 | PS | 248.33 | 137.80 | 93.34 | 323 656 |
| | | PS w/ logit | 87.78 | 43.88 | 92.80 | 322 032 |
| | | PS w/ GAM | 84.93 | 41.13 | 92.79 | 322 015 |
| | | PS w/ logit and GAM | 25.76 | 14.93 | 92.73 | 321 842 |
| | | Analytical | 1210.52 | 735.66 | 107.38 | 363 824 |
| | | Isotonic regression | 65.47 | 41.76 | 92.76 | 322 035 |
| 1 000 000 | 0.0022 | PS | 8.60 | 3.44 | 2.13 | 14 679 |
| | | PS w/ logit | 3.91 | 1.07 | 2.13 | 14 636 |
| | | PS w/ GAM | 4.75 | 1.35 | 2.13 | 14 639 |
| | | PS w/ logit and GAM | 3.68 | 1.00 | 2.13 | 14 635 |
| | | Analytical | 88.13 | 29.67 | 2.21 | 16 266 |
| | | Isotonic regression | 4.10 | 1.64 | 2.13 | 14 643 |
| | 0.0208 | PS | 75.01 | 32.61 | 20.29 | 96 127 |
| | | PS w/ logit | 30.76 | 11.22 | 20.25 | 95 732 |
| | | PS w/ GAM | 32.23 | 9.88 | 20.25 | 95 745 |
| | | PS w/ logit and GAM | 10.84 | 3.46 | 20.24 | 95 690 |
| | | Analytical | 569.12 | 238.58 | 23.48 | 108 924 |
| | | Isotonic regression | 15.25 | 6.64 | 20.24 | 95 705 |
| | 0.1109 | PS | 248.26 | 134.61 | 93.34 | 323 649 |
| | | PS w/ logit | 86.88 | 40.49 | 92.80 | 322 027 |
| | | PS w/ GAM | 84.88 | 32.82 | 92.79 | 321 992 |
| | | PS w/ logit and GAM | 17.09 | 7.90 | 92.72 | 321 831 |
| | | Analytical | 1210.52 | 735.66 | 107.38 | 363 824 |
| | | Isotonic regression | 33.62 | 20.83 | 92.73 | 321 860 |

*3.2.4 Base model is nearly perfect in expectation*

When the base model is nearly perfect in expectation, Platt's scaling with the logit transformation was generally the most effective calibration method (see Table 4). Analytical calibration was no longer perfect because of the errors in the base model's predictions, but it also did not perform nearly as poorly as it did when the base model pushed probability estimates towards 0.5 or towards extreme values. When the mean outcome success rate was 0.0022 and the calibration dataset had 100 000 observations, analytical calibration was competitive with Platt's scaling with the logit transformation as the best method. However, as the other calibration methods were given more information, either through increasing the size of the calibration dataset or through the mean outcome success rate increasing, the relative effectiveness of analytical calibration diminished. The results in Table 4 again reinforce why Platt's scaling should not be blindly used to calibrate models after undersampling; it often performed considerably worse than Platt's scaling with the logit transformation.

**Table 4** The root mean squared error (RMSE), mean absolute error (MAE), Brier score (BS), and negative logarithmic score (NLS) Platt's scaling (PS) and its variations, as well as two baseline methods. The base model used here is imperfect but nearly perfect in expectation

| Size of calibration dataset | Mean outcome success rate | Calibration method | RMSE (×10⁴) | MAE (×10⁴) | BS (×10³) | NLS |
|---|---|---|---|---|---|---|
| 100 000 | 0.0022 | PS | 7.16 | 3.64 | 2.26 | 15 369 |
| | | PS w/ logit | 6.08 | 3.53 | 2.26 | 15 363 |
| | | PS w/ GAM | 9.94 | 5.33 | 2.26 | 15 427 |
| | | PS w/ logit and GAM | 9.62 | 5.02 | 2.26 | 15 416 |
| | | Analytical | 6.20 | 3.50 | 2.26 | 15 359 |
| | | Isotonic regression | 9.97 | 5.82 | 2.26 | 15 458 |
| | 0.0208 | PS | 58.88 | 33.05 | 20.17 | 95 604 |
| | | PS w/ logit | 50.95 | 31.08 | 20.16 | 95 553 |
| | | PS w/ GAM | 52.34 | 31.48 | 20.16 | 95 559 |
| | | PS w/ logit and GAM | 51.40 | 31.12 | 20.16 | 95 553 |
| | | Analytical | 54.06 | 32.39 | 20.16 | 95 574 |
| | | Isotonic regression | 64.23 | 34.61 | 20.17 | 95 692 |
| | 0.1109 | PS | 223.06 | 151.10 | 92.36 | 321 139 |
| | | PS w/ logit | 205.99 | 144.19 | 92.28 | 320 961 |
| | | PS w/ GAM | 207.14 | 144.72 | 92.29 | 320 975 |
| | | PS w/ logit and GAM | 205.99 | 144.18 | 92.28 | 320 961 |
| | | Analytical | 213.50 | 147.77 | 92.31 | 321 041 |
| | | Isotonic regression | 215.20 | 150.11 | 92.32 | 321 256 |
| 1 000 000 | 0.0022 | PS | 6.84 | 3.54 | 2.26 | 15 368 |
| | | PS w/ logit | 5.79 | 3.34 | 2.26 | 15 360 |
| | | PS w/ GAM | 6.57 | 3.49 | 2.26 | 15 366 |
| | | PS w/ logit and GAM | 5.79 | 3.34 | 2.26 | 15 360 |
| | | Analytical | 6.20 | 3.50 | 2.26 | 15 359 |
| | | Isotonic regression | 6.89 | 3.72 | 2.26 | 15 379 |
| | 0.0208 | PS | 58.69 | 33.21 | 20.17 | 95 603 |
| | | PS w/ logit | 51.04 | 31.20 | 20.16 | 95 552 |
| | | PS w/ GAM | 51.93 | 31.36 | 20.16 | 95 553 |
| | | PS w/ logit and GAM | 51.16 | 31.19 | 20.16 | 95 552 |
| | | Analytical | 54.06 | 32.39 | 20.16 | 95 574 |
| | | Isotonic regression | 53.55 | 32.11 | 20.16 | 95 589 |
| | 0.1109 | PS | 222.87 | 150.24 | 92.36 | 321 129 |
| | | PS w/ logit | 205.43 | 143.47 | 92.28 | 320 950 |
| | | PS w/ GAM | 205.93 | 143.51 | 92.28 | 320 955 |
| | | PS w/ logit and GAM | 205.40 | 143.38 | 92.28 | 320 948 |
| | | Analytical | 213.50 | 147.77 | 92.31 | 321 041 |
| | | Isotonic regression | 207.59 | 144.62 | 92.29 | 321 004 |

### 3.3 Simulation study discussion

Our simulation study has shown that the effectiveness of different calibration methods varies substantially based on a number of factors. Although we have considered the level of class imbalance in the data and the size of the calibration dataset, our primary consideration was the performance of the base model. Our results show that this is an extremely important factor to consider. While our simulation study has shown that traditional Platt's scaling can perform very poorly in some cases, we also found that one of the four variations of Platt's scaling was the best-performing calibration method in all cases except with the perfect base model. Thus, Platt's scaling or a variation of it can be a valuable tool for calibrating models after undersampling.

Although traditional Platt's scaling performed very poorly when the base model was perfect, the variation of Platt's scaling that uses the logit transformation performed well. It could not perform as well as analytical calibration because that method perfectly adjusts for undersampling in this setting. However, analytical calibration is risky to use because it can perform very poorly with imperfect models. In addition, even with a model that is nearly perfect in expectation, Platt's scaling with the logit transformation was able to outperform it.

Traditional Platt's scaling was not effective when the base model was perfect, but it worked very well when the base model pushed probabilities towards 0.5. In this setting, the relationship between the model's outputs and the true probabilities is linear on the log odds scale, so the assumptions of the logistic regression model are met. Thus, even though Platt's scaling is unable to properly calibrate a well-calibrated base model after undersampling, it might be suitable sometimes.

When the base model pushes probabilities towards 0 or 1, Platt's scaling with both the logit transformation and a GAM tended to lead to the best-calibrated probabilities. This is one of the safest calibration choices, as it was generally in the top three regardless of the base model. Using Platt's scaling with a GAM, whether with the logit transformation or not, seems to be a reasonable choice because of its relatively strong performance and robustness to the base model. However, this approach was sometimes still considerably less effective than one of the other Platt's scaling approaches when the outcome probability and calibration dataset size were both relatively small. Similar comments can be made about isotonic regression, whose use led to identical probability estimates regardless of the base model used (excluding the results in Section 4.2.4, which involve randomness). However, although isotonic regression was robust to the base model (even more so than Platt's scaling with a GAM), its performance generally was not very good.

Our simulation study has also shown that the use of Platt's scaling can have undesirable effects on model selection. Consider the case where we fit several base models and would like to compare them so that we can choose a final model. In our simulation study, we have a perfect model, a model that pushes probabilities towards 0.5, a model that pushes probabilities towards extreme values, and a model that is imperfect but nearly perfect in expectation. If we use traditional Platt's scaling to calibrate these models, then the corresponding rows in Tables 1-4 would be used to compare the models. In our case, we would choose the model that pushes probabilities towards 0.5 as our final model, even though one of our candidate models is perfect. This could negatively affect our ability to interpret the model as we try to better understand the underlying data generating process.

Before moving on to the case study, it is important to note that relatively small differences in BS and NLS often reflect substantially different performance in terms of accurate estimation of probabilities. With access to the true probabilities, we can evaluate the calibration approaches using visualizations, as well as RMSE and MAE, allowing us to see this difference in performance. When comparing calibration methods in the case study, without access to the true probabilities, we must keep in mind that small differences in BS and NLS can reflect meaningful differences in probability estimation.

## 4. Case study

To demonstrate the impacts of the choice of calibration method in practice, we considered a version of the Cover Type dataset from the UC Irvine machine learning repository (Asuncion and Newman, 2007). In the original dataset, there were seven classes representing the types of trees in a region. Dal Pozzolo et al. (2015a) modified the dataset to create a binary classification problem. The dataset has 38 500 observations, 2747 (7.14%) of which are positive. Ten covariates are available for modelling the response.

### 4.1 Methodology

Our base model was a simple feedforward neural network that consisted of a 10-neuron input layer for batch normalization (Ioffe and Szegedy, 2015), a 10-neuron hidden layer using the rectified linear unit (ReLU) activation function (Nair and Hinton, 2010), and a one-neuron output layer using the sigmoid activation function. We trained the neural network using binary cross entropy loss with the Adam optimizer (Kingma and Ba, 2015). To prevent the model from overfitting, we used early stopping (e.g., Prechelt, 1998). If the model had not improved its performance on the validation dataset in the last 30 epochs, the training was stopped. The model was afforded at most 300 epochs.

As a result of using a neural network as our base model, we needed to partition the Cover Type dataset into four datasets: one for training the neural network; one for validating the neural network; one for training the calibration methods; and one for evaluating the modelling processes. Half of the data was allocated to training the neural network and one sixth was allocated to each of the other three purposes. When partitioning the data, we ensured that the proportion of positive cases was maintained in all four datasets. However, the training and validation datasets were then modified to adjust for class imbalance; for these datasets, we kept each negative instance with a probability of 0.1.

To evaluate and compare the calibration approaches, we computed the BS and NLS for each method on the testing dataset. We also created scatter plots comparing the probability estimates from each method.

## 4.2 Case study results

In Table 5, we show the BS and NLS for each calibration method. The Platt's scaling methods with the logit transformation performed best, followed by analytical calibration. Isotonic regression did not perform well, but traditional Platt's scaling was the clear worst choice.

**Table 5** The Brier score (BS) and negative logarithmic score (NLS) for a neural network model calibrated using Platt's scaling (PS) and its variations, as well as two baseline methods

| Calibration method | BS ($\times 10^2$) | NLS |
|---|---|---|
| PS | 2.447 | 523 |
| PS w/ logit | 2.271 | 493 |
| PS w/ GAM | 2.304 | 505 |
| PS w/ logit and GAM | 2.272 | 493 |
| Analytical | 2.281 | 500 |
| Isotonic regression | 2.321 | 519 |

Scatter plots comparing the probability estimates of each method are shown in Fig. 6. These show that traditional Platt's scaling leads to probability estimates that are very different from those obtained with the other methods. Considering its poor performance (see Table 5), it is reasonable to conclude that the estimates from this approach did a very poor job accurately reflecting the true probabilities. Using a GAM appears to have partially corrected this issue, but not entirely.

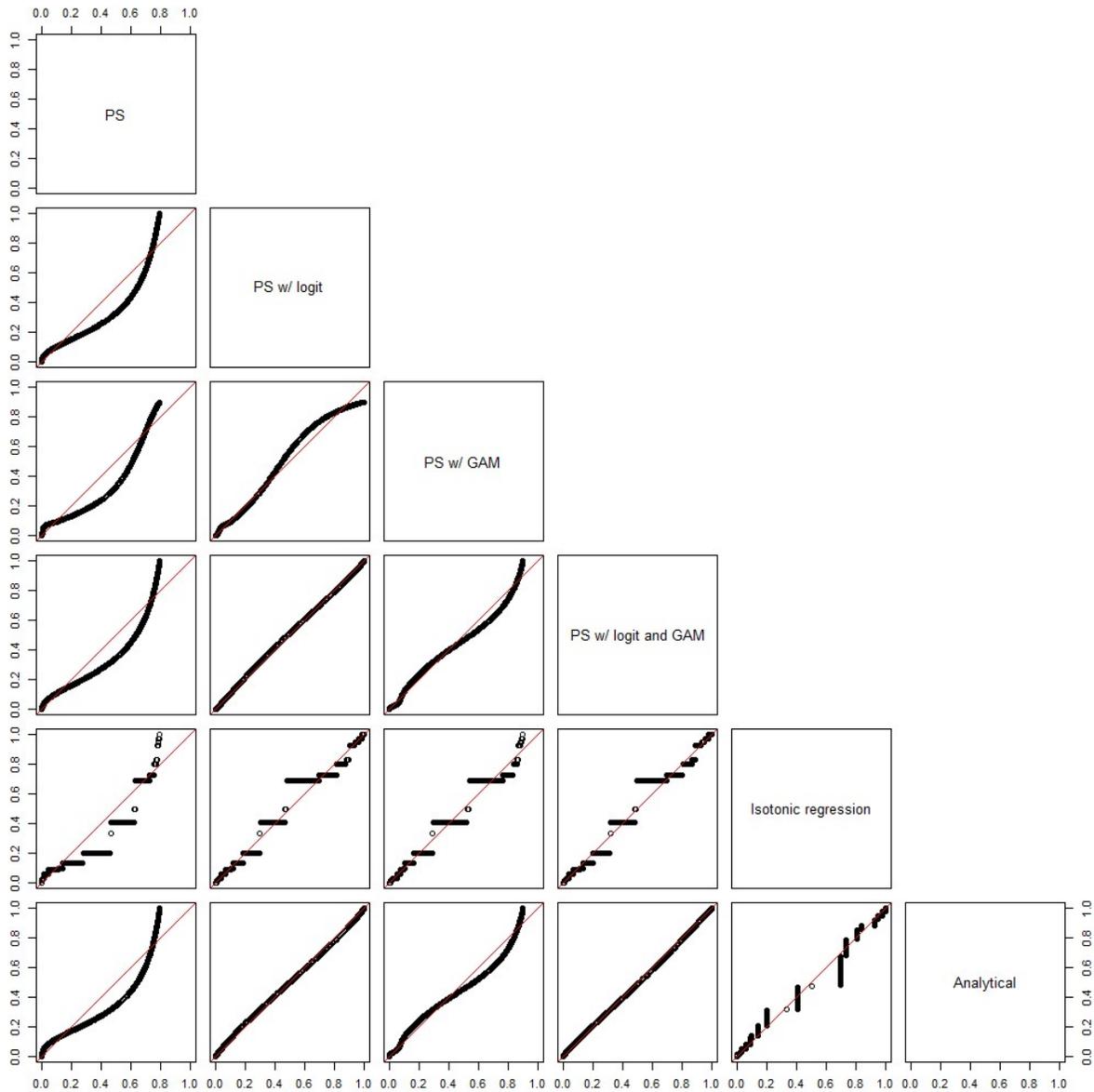

**Fig. 6** Scatter plots of the probability estimates from a neural network model calibrated using Platt's scaling (PS) and its variations, as well as two baseline methods

### 4.3 Case study discussion

Our case study has demonstrated that the choice of calibration method can have a considerable impact on the resulting probability estimates in practice. Although the differences may not seem large based on metrics like BS and NLS, the differences shown in Fig. 6 are substantial, especially for cases with relatively high probability estimates – the cases we are generally most interested in. The case study has provided even more evidence that Platt's scaling should not be blindly used to calibrate models after undersampling, as other methods can provide substantial improvements. However, it also provides evidence that variations of Platt's scaling can be valuable calibration methods after undersampling. In this case, the methods that employ the logit transformation led to the two best sets of predictions.

## 5. Conclusion

In this paper, we have shown analytically, as well as through a simulation study and a case study, that Platt's scaling is generally not a good choice for calibrating models trained on an undersampled dataset. Although it can work, Platt's scaling relies on the base model having a specific systematic error to properly calibrate the predictions. However, a modified version of Platt's scaling based on the logit of the base model's predictions is an effective calibration approach. Using a logistic GAM instead of a logistic regression model can also lead to improved calibration.

To choose a calibration method in practice, the most robust approach is to compare the different methods and then choose the best one. As done herein, a practitioner could quantitatively evaluate each method given a particular base model, and then select the best calibration method. For real data, the true probabilities are unknown, so metrics like BS or NLS would need to be used. Of the two, we recommend NLS because it can better reflect differences in the models (Benedetti, 2010), as shown in Tables 1-3. Other metrics may also be viable options, like customized metrics from the Beta family of scoring rules (Merkle and Steyvers, 2013). See Phelps and Woolford (2021b) for an example of the use of customized metrics in the context of wildfire occurrence prediction. It is also important to note that to obtain an unbiased estimate of the performance of the entire modelling procedure (i.e., with the chosen calibration method), an additional dataset is needed. Using the metrics obtained on the testing dataset (which were used to choose the calibration method) will result in a biased estimate.

This quantitative comparison is a relatively time-consuming process, so a practitioner may wish to bypass this procedure. This may be possible by critically thinking about the base model being used. In general, our results indicate that a practitioner should only use traditional Platt's scaling for calibration after undersampling if it would have been able to calibrate the base model had it been trained on the entire dataset (i.e., without undersampling) (e.g., see Böken, 2021). For example, boosted trees and random forests tend to push probability estimates towards 0.5 (e.g., Niculescu-Mizil and Caruana, 2005; Guilbert et al., 2024), so one might choose to use Platt's scaling for calibration anyway when using these models. Traditional Platt's scaling can simultaneously account for a base model pushing its estimates towards 0.5 and being miscalibrated due to undersampling, so it might be a good choice in this situation. However, this justification for using Platt's scaling is generally not given in the undersampling literature; oftentimes, the only reasoning given is miscalibration due to undersampling (e.g., Wallace and Dahabreh, 2013; Moreau et al., 2020; Peng et al., 2020; Phelps and Woolford, 2021b; Burmeister et al., 2023; Shin et al., 2023). Even when models push probability estimates towards 0.5, they cannot be expected to err in a perfectly sigmoidal fashion. Consequently, it might still be better to use Platt's scaling with a logistic GAM, especially because the GAM will converge to the logistic regression model if the assumptions of logistic regression are met.

If a practitioner cannot justify using Platt's scaling had they not used undersampling, then we recommend using the modification of Platt's scaling with the logit transformation and the GAM. This approach was relatively robust to the base model, especially when the mean outcome probabilities were larger (e.g., 0.1) or the calibration dataset was large (e.g., 1 000 000 observations). It is also supported analytically when the base model is perfect, and the GAM allows the calibration model to adjust to errors in the base model. This flexibility may prove to be very valuable in practice when errors in the base model might be more complex (e.g., asymmetrical about 0.5).

**Data availability statement:** The modified Cover Type dataset was made publicly available in a zip file in Dal Pozzolo et al. (2015a).

**Author contributions:** Nathan Phelps: Conceptualization, Formal analysis, Funding acquisition, Investigation, Methodology, Software, Validation, Visualization, Writing – original draft, Writing – review and editing; Daniel J. Lizotte: Funding acquisition, Project administration, Resources, Supervision, Writing – review and editing; Douglas G. Woolford: Funding acquisition, Project administration, Resources, Supervision, Writing – review and editing

# Appendix

For all of the simulated datasets, we generated 10 covariates. Each of these covariates followed a uniform distribution, but with different minimums and maximums. Those values are shown in Table A1.

**Table A1** The minimum and maximum value for each of the 10 covariates in the simulated datasets

| Covariate | Minimum | Maximum |
| --- | --- | --- |
| 1 | -0.4 | 0.6 |
| 2 | -0.2 | 0.8 |
| 3 | -0.4 | 1 |
| 4 | -0.1 | 0.9 |
| 5 | 0 | 5 |
| 6 | 0 | 3 |
| 7 | 1 | 4 |
| 8 | 1 | 7 |
| 9 | 1 | 3 |
| 10 | 0 | 2 |

Based on the 10 covariates, we generated the log odds of success for each observation according to Eq. A1:

$$logit(p) = \frac{\log(99)}{40}(x_1 + x_2 + x_3 + x_4 + x_5 + x_6 + x_7 + x_8 + x_9 + x_{10} + x_1 x_3 + x_2 x_5 + x_4 x_9 + x_6 x_7 + x_8 x_{10} + x_1 x_2 x_3 x_4 + x_1 x_2 x_9 x_{10}) - b \log(99) \quad \text{Eq. A1}$$

Here, $b$ is a parameter that can be used to alter the rate at which successes occur. For our simulation study, we set $b$ to 2, 1.5, and 1.1. Undoing the logit operation yields the success probabilities, which were used to simulate the outcomes in the datasets.